\documentclass[preprint]{revtex4}
\usepackage{epsfig}

\begin{document}

\title{GL(2,R) dualities in generalised Z(2) gauge theories and Ising models}
	\author{Nelson Yokomizo}
\email{yokomizo@fma.if.usp.br}
\author{Paulo Teotonio-Sobrinho}
\email{teotonio@fma.if.usp.br}
\affiliation{Instituto de F\'{i}sica, Universidade de S\~{a}o Paulo, C.P. 66318, 05315-970 S\~{a}o Paulo-SP, Brazil}
\date{\today}

\begin{abstract}
{We study a class of duality transformations in generalised Z(2) gauge theories and Ising models on two- and three-dimensional compact lattices. The theories are interpreted algebraically in terms of the structure constants of a bidimensional vector space $\mathcal{H}$ with algebra and coalgebra structures, and it is shown that for any change of basis in $\mathcal{H}$ there is a related symmetry between such models. The classical Kramers and Wannier dualities are described as special cases of these transformations. We derive explicit expressions for the relation between partition functions on general finite triangulations for these cases, extending results known for square and cubic lattices in the thermodynamical limit. A class of symmetry transformations in which the gauge coupling changes continuously is also studied in two dimensions.}

\end{abstract}

\maketitle

\section{Introduction}

Duality relations \cite{wegner} are symmetries which relate partition functions of classical spin models. The prototype for these symmetries, and its most famous example, is the well-known Kramers and Wannier duality in the two-dimensional (2D) Ising model \cite{kramers-wannier}. This symmetry relates the partition function of a 2D Ising model at temperature $T$ to the partition function of another 2D Ising model at temperature $T^\star$, where $T^\star$ is a decreasing function of $T$, and allowed for the first exact determination of the critical point of the model, before its exact solution by Onsager \cite{onsager}. If it is known that the critical point is unique, then it must be at $T_c = T_c^\star$, and can be determined by the duality relation. Besides its interest in the determination of critical points \cite{kramers-wannier,bhanot}, dualities are also used to map properties of well-studied models to less developed ones \cite{map,map2}. For instance, it is known that in three dimensions Z(2) lattice gauge theory is dual to the Ising model. Thus all work developed in understanding the latter can be translated via duality relations to bring information about the former \cite{map3d}. In this paper we explore duality relations in Z(2) lattice gauge theories and Ising models in two and three dimensions making use of an algebraic reformulation of spin models we developed for the case of the Z(2) pure gauge theory in an earlier paper \cite{reform}.

The Kramers-Wannier dualities can be briefly stated as follows \cite{wegner}. For each dimension $d$ and integer $n$, $1 \leq d \leq n$, there is one $d$-dimensional generalised Ising model denoted $M_{dn}$, whose local variables are Ising spins $\sigma_a=\pm 1$. The index $n$ indicates where the spins are situated. For spins lying on vertices, links and faces we have $n=1$, $2$ and $3$, respectively. The terms in the hamiltonian are products of spins lying on the boundary of links for $n=1$, of faces for $n=2$, etc. Therefore, $M_{d1}$ describes the usual $d$-dimensional Ising model, with two-spin interactions and Ising spins at vertices. $M_{d2}$ is the Z(2) pure gauge theory in $d$ dimensions, with spins lying on links and plaquette terms in the hamiltonian. An external magnetic field $h$ can be added to any $M_{dn}$. If that is the case, then $M_{d1}$ becomes the Ising model with an external magnetic field, and $M_{d2}$ describes the Z(2) gauge theory coupled to a Higgs matter field. There are two kinds of duality relations. In the absence of magnetic field, the models $M_{dn}$ and $M_{d\,d-n}$ are dual. In the presence of a magnetic field, the models $M_{dn}$ and $M_{d\,d-n+1}$ are dual. For the simplest cases, one finds the following. In two dimensions, the Ising model is self-dual in the absence of magnetic field, and dual to the Z(2) Higgs-gauge theory in the presence of magnetic field. In three dimensions, the Ising model is dual to Z(2) pure gauge theory, and the Z(2) Higgs-gauge theory is self-dual.

We extend these dualities in two directions. First, we consider more general lattices than the square and cubic lattices traditionally used. Second, we describe them as special cases of a larger class of GL(2,$\mathbf{R}$) symmetries acting on generalised Z(2) lattice gauge theories herein defined. A mathematical formalism borrowed from topological field theories is used to develop an algebraic description of Z(2) gauge-Ising models, following the procedure introduced in \cite{reform} for the case of the Z(2) pure gauge theory on three-dimensional triangulations. This leads us to associate a bidimensional vector space $\mathcal{H}$ with algebra and coalgebra structures to each model under study. If the vector basis of $\mathcal{H}$ is changed, a lattice model equivalent to the original one is obtained. We study two specific changes of basis of $\mathcal{H}$. One transformation leads to explicit expressions for the Kramers-Wannier dualities on arbitrary lattices. The cases of triangulations and regular lattices in two and three dimensions are worked out in detail. Usually, one needs to study these dualities case-by-case, applying sequences of combinatorial moves as the decoration transformation or the star square transformation \cite{wegner} in a manner suited for each lattice under consideration. In the algebraic formalism, all these operations are shown to be contained in a single change of basis of $\mathcal{H}$. In addition, we study an one-parameter class of symmetry transformations for which the gauge coupling constant changes continuously in the two-dimensional generalised Z(2) gauge theories as another application of the general formalism.

This paper is organised as follows. In Section 2 we give a brief description of Z(2) gauge theories and Ising models in $d$-dimensions, mainly to fix notation. In Section 3 we introduce our algebraic reformulation of the Z(2) lattice gauge theory. The cases of the pure and Higgs-gauge theories are considered, for triangulations and hypercubic lattices. In Section 4 we define the generalised gauge theories studied in this paper and prove that for any change of basis in $\mathcal{H}$ there is a related duality transformation between these models. We work out many examples in two and three dimensions. In particular, we describe the classical Kramers-Wannier dualities as special instances of our formalism. In Section 5 we study in two dimensions the one-parameter class of symmetry transformations for which the gauge coupling changes continuously. Our conclusions are collected in Section 6, together with final remarks and the discussion of possible developments.

\section{The models}

The $d$-dimensional Ising model is determined for any $d$-dimensional lattice $L$ by two coupling constants: the inverse temperature $\beta_I$, and the external magnetic field $m$. The local variables are Ising spins $\sigma_v=(-1)^v,v=0,1$, assigned to the vertices $v$ of $L$. In terms of these, the hamiltonian reads
\begin{equation}
H_I=\beta_I \sum_l \sigma_x(l) \, \sigma_y(l)+ m \sum_v \sigma_{v} \, \textrm{,}
\end{equation}
where $\sigma_x(l)$ and $\sigma_y(l)$ are the spins at the two ends of the link $l$, and $\sigma_v$ is the spin at the vertice $v$. The first sum runs over all links, and the second over all vertices of $L$. The partition function is
\[
Z_I = \sum_{ \left\{ \sigma_v \right\} } \exp(H_I) \, \textrm{,}
\]
where $\left\{ \sigma_v \right\}$ denotes the set of all spin configurations over the vertices of $L$. This partition function can be rewritten as
\begin{equation}
Z_I = \sum_{ \left\{ \sigma_v \right\} } \prod_l W_I(l) \prod_v W_I(v) \, \textrm{,} \label{eq:Zising}
\end{equation}
where we have introduced the local Boltzmann weights
\begin{equation}
\begin{array}{c}
W_I(l) = \exp[\beta_I \, \sigma_x(l) \, \sigma_y(l)] \, \textrm{,} \\
W_I(v) = \exp(m \sigma_v) \, \textrm{,} \label{eq:weights-ising}
\end{array}
\end{equation}
situated at the links and vertices of $L$, respectively. The case $m=0$ describes the Ising model in the absence of an external magnetic field, in which case $W_I(v)=1$ and can be ignored in Eq. \ref{eq:Zising}. If $m \neq 0$, then we say that there is an external magnetic field.

The Z(2) lattice gauge theory is also determined by two coupling constants, the gauge coupling $\beta_g$, and the Higgs coupling $h$. The gauge variables are elements of the discrete group Z(2), i.e. Ising spins $\sigma_a=(-1)^a,\,a=0,1$. There is one spin-gauge variable $\sigma_a$ at each link $a$ of the lattice $L$. The action is
\begin{equation}
S_g=\beta_g \sum_f \prod_a \sigma_a(f) + h \sum_l \sigma_{l} \, \textrm{,} \label{eq:z2action}
\end{equation}
where $\sigma_a(f)$ is the spin at the link $a$ on the boundary of the face $f$, and the product runs over all links on $\partial f$. The first sum runs over all faces, and the second over all links of $L$. When $L$ is a triangulation, all faces are triangular and the first term describes a three-spin interaction, $\sigma_a(f) \sigma_b(f) \sigma_c(f)$. If $L$ is a cubic lattice, then this term gives a four-spin interaction. In general lattices, the number of spins in the product is variable, equal in each face to the number of links on its boundary.

The partition function of Z(2) gauge theory is given by the usual expression
\[
Z_g = \sum_{ \left\{ \sigma_l \right\} } \exp(S_g) \, \textrm{,}
\]
where the sum runs over all spin configurations $\left\{ \sigma_a \right\}$ over the links of $L$. This partition function can be rewritten as
\begin{equation}
Z_g = \sum_{ \left\{ \sigma_l \right\} } \prod_f W_g(f) \prod_l W_g(l) \, \textrm{,} \label{eq:Zgauge}
\end{equation}
where we have introduced the local Boltzmann weights
\begin{equation}
\begin{array}{c}
W_g(f) = \exp[\beta_g \, \prod_a \sigma_a(f)] \, \textrm{,} \\
W_g(l) = \exp(h \, \sigma_l) \, {,} \label{eq:weightsZ2}
\end{array}
\end{equation}
situated at the faces and links of $L$, respectively. When $h=0$, we have Z(2) pure gauge theory. In this case $W_g(l)=1$ and can be ignored in Eq. \ref{eq:Zgauge}.
If $h\neq 0$, then we say that Z(2) gauge theory is coupled to a Higgs matter field.

The definitions make sense for any dimension $n$. When a specific dimension is considered, we write $H_I^{(2)}$, $Z_g^{(3)}$, etc.

\section{Algebraic reformulation of Z(2) gauge theory}
In this section we present an algebraic reformulation of Z(2) lattice gauge theory in which the local Boltzmann weights and the partition function are described as coefficients of certain tensors in a suitable vector space $\mathcal{H}$. These tensors endow $\mathcal{H}$ with algebra and coalgebra structures. We first give a characterisation of $\mathcal{H}$ by itself, in a mathematical independent way, and show how a lattice field theory with face and link local weights can be constructed from the knowledge of the structure constants of $\mathcal{H}$. Next we adjust the free parameters of $\mathcal{H}$ so that the weights correspond to those of Z(2) lattice gauge theory. We consider the cases of pure and Higgs-gauge theory on triangulations and hypercubic lattices.

\subsection{Structure tensors and weights}

Let us describe the algebraic structure $\mathcal{H}$. We define $\mathcal{H}$ as a bidimensional vector space with basis $\mathcal{B}=\{\phi_0,\phi_1\}$, and dual basis $\mathcal{B}^\star=\{\psi^0,\psi^1\}$, with $\psi^a(\phi_b)=\delta^a_b$. There is a linear product of vectors defined in $\mathcal{H}$, described by a product tensor $M_{ab}^c$, for which
\begin{equation}
\left( u \cdot v \right)^c = M_{ab}^c u^a v^b, \qquad \forall u,v \in \mathcal{H} \, \textrm{.}
\end{equation}
In the basis $\mathcal{B}$, the product of basis vectors read
\begin{equation}
\begin{array}{ccc}
\phi_0 \cdot \phi_0 = &\phi_1 \cdot \phi_1 = &\rho^{-1} \cosh(x) \; \phi_0 + \rho^{-1} \sinh(x) \; \phi_1 \, {,} \\
\phi_0 \cdot \phi_1 = &\phi_1 \cdot \phi_0 = &\rho^{-1} \sinh(x) \; \phi_0 + \rho^{-1} \cosh(x) \; \phi_1 \, \textrm{.} \label{eq:alg}
\end{array}
\end{equation} 
The algebra so defined is associative, and has an unity $\iota = \rho \, \cosh(x) \, \phi_0 - \rho \, \sinh(x) \, \phi_1$. The parameters $\rho$ and $x$ will be given as functions of the gauge coupling $\beta_g$ for each lattice $L$ under consideration. There is also a coproduct defined in $\mathcal{H}$, described by a coproduct tensor $\Delta_a^{bc}$, for which
\begin{equation}
\left[ \Delta(u) \right]^{bc} = \Delta_a^{bc} u^a, \; \forall u \in \mathcal{H} \, \textrm{.}
\end{equation}
We define $\Delta$ by its action on the basis vectors of $\mathcal{B}$, which is given by
\begin{equation}
\Delta(\phi_a)= K_a \, \phi_a \otimes \phi_a \, \textrm{,} \label{eq:coprod}
\end{equation}
where $K_a$ is a constant which will be determined as a function of the Higgs coupling $h$ for each lattice $L$. In this basis, the coproduct tensor coefficients read
\begin{equation}
\Delta_a^{bc} =  K_a \, \delta_{ab} \, \delta_{ac} \, \textrm{.} \label{eq:coalg}
\end{equation}
The coalgebra so defined is coassociative, and has a counity $\varepsilon=K_0^{-1} \psi^0 + K_1^{-1} \psi^1$. 

We now proceed to show how the product and coproduct tensors $M_{ab}^c$ and $\Delta_a^{bc}$ can be used in order to define the local Boltzmann weights of a lattice field theory. For that purpose we follow prescriptions introduced in \cite{kuperberg} in the context of topological field theories, and adapted to the case of lattice gauge theories in \cite{reform}. These prescriptions allow one to define local weights at faces and links on any lattice given as the gluing of polygonal faces along common links, as triangulations and hypercubic lattices, for example. It is sufficient for this that $\mathcal{H}$ is associative and coassociative. It was proved in \cite{kuperberg} that whenever $\mathcal{H}$ is a Hopf algebra, the resulting theory is a topological field theory. Here we consider an algebra which is not Hopf, and show that it leads to Z(2) lattice gauge theory in a variety of conditions \footnote{We have not defined an antipode operator $S: \mathcal{H} \mapsto \mathcal{H}$, but we will admit that $S_a^b=\delta_a^b$, where $\delta$ is the Kronecker delta.}. The reformulation of a lattice gauge theory in algebraic terms is not trivial, and allows for the use of new mathematical tools in its study, as we will show for the case of duality relations.

\begin{figure}
\center
\includegraphics[scale=0.55]{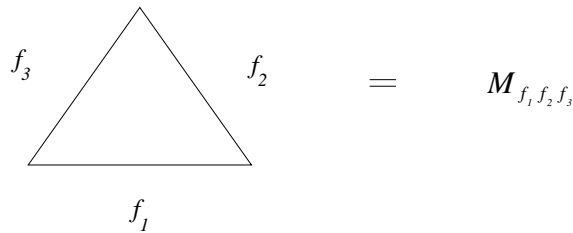}
\caption{A triangular face $f$ with configurations $f_i=0,1$ on its boundary, describing spin-gauge variables $\sigma_{f_i}=(-1)^{f_i}$. The corresponding local Boltzmann weight is given by $M_{f_1 f_2 f_3}$.}
\label{fig:faceweight}
\end{figure}

The prescriptions we use consist of a set of formulae in which the structure tensors $M_{ab}^c$ and $\Delta_a^{bc}$ are used to construct higher order tensors, whose coefficients are identified with local weights of a lattice theory associated to $\mathcal{H}$. The lattice $L$ is always supposed to be a triangulation or a hypercubic lattice, for definiteness. Let us first consider the face weights. These are determined by the algebra structure of $\mathcal{H}$. We define the covariant tensors $M_{a b \dots c}$ as
\begin{equation}
M_{a b \dots c} = T(\phi_{a} \, \phi_{b} \cdots \phi_{c}) \, \textrm{,} \label{eq:faceweights}
\end{equation}
where
\begin{equation}
T(v)=M_{ab}^b \, v^a, \; \forall v \in \mathcal{H} \, \textrm{.}
\end{equation}
The covariant vector $T_a=M_{ab}^b$ is called the trace. The tensor $M_{a b \dots c}$ so defined is cyclically symmetric whenever the algebra $\mathcal{H}$ is associative \cite{kuperberg}. Now consider a face $f$ of $L$, and let $n$ be the number of links on its boundary. In addition, let there be spins $\sigma_{f_i}$ at the links on its boundary, where $f_i=0,1$. Then it is assigned to this face a local Boltzmann weight given by $M_{f_1 f_2 \dots f_n}$ (see Fig. \ref{fig:faceweight}). Its value depends only on the structure constants $M_{ab}^c$ of $\mathcal{H}$, and on the spin configurations at the boundary of $f$. 

The link weights are determined in a similar fashion by the coalgebra structure of $\mathcal{H}$. In analogy with Eq. \ref{eq:faceweights}, we define contravariant tensors
\begin{equation}
\Delta^{a b \dots c} = C(\psi^{a} \psi^{b} \dots \psi^{c}) \, \textrm{,} \label{eq:linkweights}
\end{equation}
where
\begin{equation}
C(\omega)=\Delta^{ab}_{a} \omega_b, \forall \omega \in \mathcal{H} \, \textrm{,}
\end{equation}
and the contravariant vector $C^a=\Delta^{ba}_{b}$ is called the cotrace. For the coproduct given in Eq. \ref{eq:coprod}, it follows that
\begin{equation}
\Delta^{a_1 \, a_2 \dots a_{m-1} \, a_m} = K_a^m \, \delta_{a_1,a} \delta_{a_2,a} \dots \delta_{a_m,a} \, \textrm{,} \label{eq:linkconfig}
\end{equation}
where the $\delta's$ are Kronecker deltas, and $a=0,1$. Let $l$ be a link of $L$, and $m$ be the number of faces meeting at $l$. Then we assign a tensor $\Delta^{a_1 a_2 \dots a_m}$ with $m$ indices to this link. According to Eq. \ref{eq:linkconfig}, the tensor coefficients vanish unless all its indices $a_i$ are equal to some fixed $a=0,1$. We let this index describe the local spin configuration $\sigma_a=(-1)^a$ at the link. The corresponding link weight is given by the tensor coefficient $\Delta^{a a \dots a}=K_a^m$.

The partition function is defined as a sum over configurations, as usual. There is one local weight $M_{f_1 f_2 \dots f_n}$ for each face $f$, and one local weight $\Delta^{a_l a_l \dots a_l}$ for each link $l$ of $L$. For a given spin configuration $\left\{ \sigma_a \right\}$, there corresponds a statistical weight
\begin{equation}
W = \prod_f M_{f_1 f_2 \dots f_n} \prod_l \Delta^{a_l \, a_l \dots a_l} \, \textrm{,} \label{eq:globalweight}
\end{equation}
where the first product runs over all faces, and the second over all links of $L$. The partition function $Z$ is the sum of $W$ over all spin configurations,
\begin{equation}
Z = \sum_{\{ \sigma_a \} } \prod_f M_{f_1 f_2 \dots f_n} \prod_l \Delta^{a_l \, a_l \dots a_l} \, \textrm{.} \label{eq:Z-kup}
\end{equation}
The expression in Eq. \ref{eq:Z-kup} completes our description of the lattice field theory associated to the algebra $\mathcal{H}$. The partition function $Z$ depends on the lattice $L$, and on the parameters $\beta$, $x$ and $K_a$ which determine the structure constants of $\mathcal{H}$. In the next sections we will show how these parameters can be adjusted to give Z(2) theory local weights, both for the case of triangulations and hypercubic lattices. But before we proceed, we discuss the algebraic meaning of the partition function $Z$. Our purpose is to prove that $Z$ is a scalar of $\mathcal{H}$. This is the central result involved in the proof of the symmetry transformations described in this paper.

Let us first rewrite the sum in Eq. \ref{eq:Z-kup} in a more symmetric form. Note that the indices of the face weights may assume any set of conceivable values inside the sum, while the indices of the link weights appear only as repetitions, $a_1=a_2= \cdots = a_l$. This assymetry can be removed. From Eq. \ref{eq:linkconfig}, we know that the tensor coefficients $\Delta^{a_1 a_2 \dots a_{n-1} a_m}$ are zero whenever any two of its indices are distinct. Thus we can add terms with local weights given by such coefficients to the sum in Eq. \ref{eq:Z-kup}, without changing the value of $Z$. In this way a new partition function $Z^\prime$ is defined, which differs from $Z$ only by a series of vanishing terms. This partition function can be made much more symmetric than the original, as we want. The new models thus obtained are the generalised spin-gauge models involved in the duality relations we will discuss later. We now turn to their definition.

\begin{figure}
\center
\includegraphics[scale=0.55]{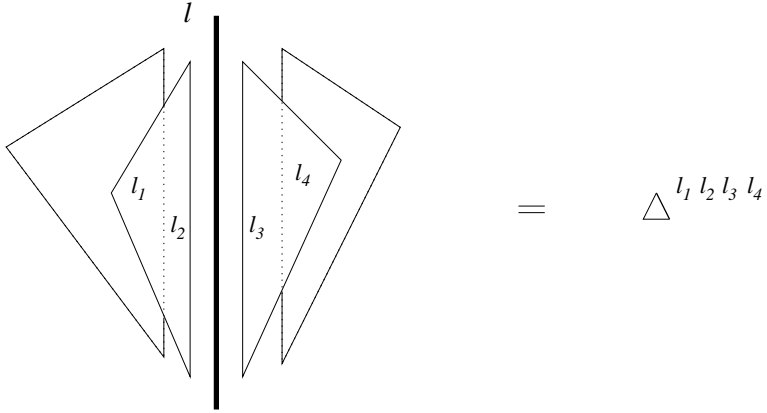}
\caption{A link $l$ shared by four faces. There are spin-gauge configurations $\sigma_{l_i}=(-1)^{l_i}$ at the face sides glued to the link. The corresponding local Boltzmann weight is $\Delta^{l_1 \, l_2 \, l_3 \, l_4}$.}
\label{fig:linkweight}
\end{figure}

Consider a particular link $l$ in a lattice $L$, and let $m$ be the number of faces meeting at the link. Instead of assigning a single spin variable $a_l$ to the link, as usual, let us attach $m$ spins to it, one for each gluing of a face to the link. One may think that the spins are attached to the boundaries of the faces, and meet at the link when the faces are glued into it (see Fig. \ref{fig:linkweight}). We write the spin variables as $\sigma_{l_i}=(-1)^{l_i}$, $l_i=0,1$. To each link let there correspond a local Boltzmann weight given by the tensor coefficients $\Delta^{l_1 l_2 \dots l_m}$, where this tensor is defined by Eq. \ref{eq:linkweights}. The face weights are $M_{f_1 f_2 \dots f_n}$, as before. Then the partition function is
\begin{equation}
Z^\prime = \sum_{\{\sigma^\prime_a\}} \prod_f M_{f_1 f_2 \dots f_n} \prod_l \Delta^{l_1 l_2 \dots l_m} \, \textrm{,} \label{eq:kup-weights}
\end{equation}
where the sum runs over all configurations $\{ \sigma_a^\prime \}$ of the new theory. In this expression, links and faces are treated on the same footing. But from Eq. \ref{eq:linkconfig}, the link weights $\Delta^{l_1 l_2 \dots l_m}$ are nonzero only if all its indices are equal to some $a_l$, and so the sum in Eq. \ref{eq:kup-weights} reduces to the one in Eq. \ref{eq:Z-kup}. Thus we conclude that $Z^\prime=Z$. 

The advantage of this formalism is that the sum in Eq. \ref{eq:kup-weights} now consists of a series of contractions of indices. Note that each spin configuration appears as an upper index in some face weight, and as a lower index in some link weight. To sum over its two values is the same as contracting this pair of indices. Furthermore, the sum over all spin configurations $\{\sigma^\prime_a\}$ is equivalent to a contraction of all indices in the tensor inside the sum symbol, i.e., the outer product of all weights $M_{f_1 f_2 \dots f_n}$ and $\Delta^{l_1 l_2 \dots l_m}$ over the lattice. The result must be a scalar, and is the partition function $Z^\prime$. Thus $Z$ is a scalar invariant of $\mathcal{H}$.

\subsection{Triangulations} \label{subsec:triang}

Now we begin to study the theories defined by the algebra $\mathcal{H}$ for specific values of the free parameters $\rho$, $x$ and $K_a$. We first let the lattice $L$ be a triangulation, and show which values these parameters must assume in order to describe Z(2) lattice gauge theory. For triangulations, the algebraic formalism can describe the pure gauge theory in any dimension, and the Higgs-gauge theory in two dimensions. In higher dimensions, a modified Higgs-gauge theory is also considered.

\paragraph*{Pure gauge theory.}

In a triangulation $L$ all faces are triangular, and so the face weights are coefficients of the three-indexed tensor $M_{f_1 f_2 f_3}$. For simplicity, we write the spin configurations on the boundary of a face $f$ as $a, b, c=0,1$. The corresponding face weight is $M_{abc}$. For Z(2) lattice gauge theory, from Eq. \ref{eq:weightsZ2} we must have
\begin{equation}
M_{abc} = \exp[(-1)^{a+b+c} \, \beta_g] \, \textrm{.}
\end{equation}
These weights are obtained from $\mathcal{H}$ when the parameters $x$ and $\rho$ satisfy
\begin{equation}
\begin{array}{ccc}
e^{-2 \beta_g}  & = &\tanh(3x) \, \textrm{,} \\
\rho^6 & = & 2 \sinh(6x) \, \textrm{,} \label{eq:3f-fix}
\end{array}
\end{equation}
as can be checked. These relations fix completely the algebraic structure constants of $\mathcal{H}$, and give the correct Z(2) face weights whenever triangular faces are considered.

The coproduct of $\mathcal{H}$ is related to the link weights. For the Z(2) pure gauge theory, all link weights $\Delta^{a_l a_l \cdots a_l}$ must be equal to $1$. This is equivalent to setting $K_a=1$, as seen from Eq. \ref{eq:linkconfig}. Thus, the Z(2) pure gauge theory is described by the coproduct
\begin{equation}
\Delta(\phi_a) = \phi_a \otimes \phi_a \, \textrm{.} \label{eq:cop-pure}
\end{equation}
The validity of this formula is not restricted to triangulations. It gives the correct link weights of the Z(2) pure gauge theory for any lattice $L$ and dimensionality $d$.

\paragraph*{Higgs-gauge theory.}

Now let $h\neq0$. First let us consider the case of two-dimensional triangulations. Then any link $l$ is shared by exactly two faces, and the corresponding link weight is $\Delta^{a_l a_l}$. Thus, from Eq. \ref{eq:weightsZ2}, we must have $\Delta^{a_l a_l} = \exp[(-1)^{a_l} h]$. From Eq. \ref{eq:linkconfig}, this condition leads to $K_0=e^{h/2}$ and $K_1=e^{-h/2}$, which is equivalent to
\begin{equation}
\begin{array}{ccc}
\Delta(\phi_0) & = & e^{h/2} \, \phi_0 \otimes \phi_0 \, \textrm{,}\\
\Delta(\phi_1) & = & e^{-h/2} \, \phi_1 \otimes \phi_1 \, \textrm{.} \label{eq:cop-2d}
\end{array}
\end{equation}
This is the coproduct for Z(2) gauge-Higgs theory in two dimensions. In higher dimensions, there is no choice of $K_a$ which leads to the correct link weights of Z(2) gauge-Higgs theory on triangulations. In general, the number $m$ of faces meeting at a link $l$ is variable in a $d$-dimensional lattice, for $d > 2$. Therefore, one is naturally led to the equation $K_a^m=\exp[(-1)^a h], \forall m \in \mathbf{N}$, which obviously has no solution for $h\neq 0$. On the other hand, a modified theory can be obtained if the coproduct of Eq. \ref{eq:cop-2d} is applied for $d>2$. That is, put 
\begin{equation}
\begin{array}{ccc}
\Delta(\phi_0) & = & e^{h} \, \phi_0 \otimes \phi_0 \, \textrm{,}\\
\Delta(\phi_1) & = & e^{-h} \, \phi_1 \otimes \phi_1 \, \textrm{.} \label{eq:cop-mod}
\end{array}
\end{equation}
This leads to the link weights
\begin{equation}
\Delta^{a_1 \, a_2 \dots a_{m-1} \, a_m} = \exp\left[(-1)^a m h \right] \delta_{a_1,a} \delta_{a_2,a} \dots \delta_{a_m,a} \, \textrm{,}
\end{equation}
i.e., to a Z(2) gauge-Higgs theory with a variable Higgs coupling $h_l$, whose strength is proportional to the number $m$ of faces meeting at the link $l$, $h_l = m h$. When the lattice is regular, one obtains a Higgs-gauge theory with Higgs coupling $m h$.

\subsection{Hypercubic lattices} \label{subsec:cubic}

In hypercubic lattices, both the pure and the Higgs-gauge theory admit an algebraic description for any dimensionality $d$. The structure constants of the product of vectors in $\mathcal{H}$ is related to the face weights of the model. All faces are square in a hypercubic lattice. Therefore, the corresponding weights are coefficients of the four-indexed tensor $M_{f_1 f_2 f_3 f_4}$. We write the configurations at the boundary of a square face as $a,b,c,d=0,1$, and the corresponding face weight as $M_{abcd}$. For Z(2) lattice gauge theory, we must have
\begin{equation}
M_{abcd} = \exp[(-1)^{a+b+c+d} \, \beta_g] \, \textrm{.}
\end{equation}
These weights follow from the relations
\begin{equation}
\begin{array}{ccc}
e^{-2 \beta_g} & = &\tanh(4x) \, \textrm{,} \\
\rho^8 & = & 2 \sinh(8x) \, \textrm{,} \label{eq:4f-fix}
\end{array}
\end{equation}
which play the role of Eq. \ref{eq:3f-fix} for the case of square faces, determining the dependence of $\rho$ and $x$ in the gauge coupling $\beta_g$.

The coproduct of $\mathcal{H}$ is determined by the link weights. In two dimensions, nothing changes in relation to the case of triangulations. The Z(2) pure and gauge-Higgs theories are described by the coproducts of Eqs. \ref{eq:cop-pure} and \ref{eq:cop-2d}, respectively, by the same arguments. In higher dimensions, hypercubic lattices offer the advantage over triangulations that the number $m$ of faces meeting at a link is now constant, due to the regularity of the lattice. For a $d$-dimensional hypercubic lattice, we have $m=2^{d-1}$. It follows that an algebraic description is then available for the Higgs-gauge theory. From Eqs. \ref{eq:weightsZ2} and \ref{eq:linkconfig}, we just need to set $K_0 = \exp(h/2^{d-1})$ and $K_1 = \exp(-h/2^{d-1})$. For instance, for cubic lattices the coproduct is
\begin{equation}
\begin{array}{ccc}
\Delta(\phi_0) & = & e^{h/4} \, \phi_0 \otimes \phi_0  \, \textrm{,}\\
\Delta(\phi_1) & = & e^{-h/4} \, \phi_1 \otimes \phi_1 \, \textrm{,} \label{eq:cop-3d}
\end{array}
\end{equation}
and for $d=2$ we recover the result of Eq. \ref{eq:cop-2d}.

\section{The duality relations}

\subsection{General algebraic symmetries}

To the fact that the partition function $Z$ is a scalar invariant of $\mathcal{H}$, there corresponds a class of symmetries among the models we described algebraically in the last section. These comprise a large class of transformations on $\mathcal{H}$. Let us recall that the local weights $M_{a_1 a_2 \dots a_n}$ and $\Delta^{a_1 a_2 \dots a_m}$ are tensor coefficients, and therefore not basis independent. Instead, if one chooses a new basis $\mathcal{B}^\prime=\{\xi_0,\xi_1 \}$, where $\xi_a=E_a^b \, \phi_b$, and $E$ is a nonsingular $2$-by-$2$ matrix, then the weights transform according to the usual formulae of tensor algebra,
\begin{equation}
\begin{array}{ccc}
\Delta^{^\prime \; l_1 \, l_2 \cdots l_m} & = & D^{l_1}_{x_1} \, D^{l_2}_{x_2} \cdots D^{l_m}_{x_m} \, \Delta^{x_1 \, x_2 \cdots x_m} \; \textrm{,}\\
M^{^\prime}_{f_1 \, f_2 \cdots f_n} & = & E_{f_1}^{x_1} \, E_{f_2}^{x_2} \cdots E_{f_n}^{x_n} \, M_{x_1 \, x_2 \cdots x_n} \; \textrm{,}
\end{array}
\end{equation}
where $D=E^{-1}$. On the other hand, as the partition function is a scalar, its value is not affected by the transformation. In short, the change of basis gives new face and link $M^{^\prime}_{f_1 \, f_2 \cdots f_n}$ and $\Delta^{^\prime \; l_1 \, l_2 \cdots l_m}$, which define a lattice theory which can be very different from the original one, but whose partition function has the same value as that of the former. For any element of GL(2,$\mathbf{R}$), there is a symmetry of this kind.

A simple example of these transformations is given by rescaling the basis vectors of $\mathcal{B}$, i.e., by letting $\xi_a = s_a \, \phi_a$, where the $s_i$ are real parameters. The local weights change as
\begin{equation}
\begin{array}{ccc}
\Delta^{^\prime \; l_1 \, l_2 \cdots l_m} & = & s_{l_1}^{-1} \, s_{l_2}^{-1} \cdots s_{l_m}^{-1} \, \Delta^{l_1 \, l_2 \cdots l_m} \; \textrm{,}\\
M^{^\prime}_{f_1 \, f_2 \cdots f_n} & = & s_{f_1} \, s_{f_2} \cdots s_{f_n} \, M_{f_1 \, f_2 \cdots f_n} \; \textrm{.}
\end{array}
\end{equation}
The interpretation of the above equations is the following. Let $\sigma_{f_i}$ be the spins at the boundary of a face $f$. Then the transformation inserts a factor $s_{f_1}$ in the corresponding face weight, for each $\sigma_{f_i}$. To compensate, an additional factor $s_{f_i}^{-1}$ is inserted in the weight of the link where $\sigma_{f_i}$ is glued. It was shown in \cite{reform} that such transformation can be used in the study of the low-temperature limit of Z(2) pure gauge theory, and leads to a clear interpretation of this regime as a topological field theory.

Hereafter we consider new instances of such transformations. We take as the starting point the algebra and coalgebra structures defined in Sections \ref{subsec:triang} and \ref{subsec:cubic} on the bidimensional vector space $\mathcal{H}$, which give Z(2) lattice gauge theory in a variety of conditions, and study the action of specific elements of GL(2,$\mathbf{R}$) on the local weights of these theories. The models obtained by these transformations may be very distinct. For instance, we will show that a transformation $E$ exists for which the face weights in the new basis $\mathcal{B}^\prime$ have the form
\begin{equation}
M^{^\prime}_{f_1 \, f_2 \cdots f_n} = \delta_{f_1 a_f} \delta_{f_2 a_f} \cdots \delta_{f_n a_f} \, \textrm{,} \label{eq:dual-latt}
\end{equation}
where the $\delta$'s are Kronecker deltas, and $a_f=0,1$. In this case, the statistical weight of a given face $f$ is nonzero only if all configurations $f_i$ at its boundary are equal to some fixed value $a_f$. One may think of such configuration as assigned to the face itself. Thus in the new model the spin variables are situated at faces. The link weights are $\Delta^{^\prime \; l_1 \, l_2 \cdots l_m}$, where the variables $l_j$ describe the configurations at the faces meeting at the link. If in the original theory there were configurations at links and local weights at faces, then we have a transformation which a model on the lattice $L$ to a model on its dual lattice $L^\star$. This is the case of the dualities depicted in \cite{wegner}, and we will show that they can be obtained as special cases of our transformations. This is done in Sections \ref{subsec:2dweg} and \ref{subsec:3dweg}. The algebraic reformulation of these dualities allows for their study on arbitrary triangulations, and leads to explicit expressions for the relation between partition functions on any finite triangulation, extending results known for square and cubic lattices.

A different sort of symmetry is considered in Section \ref{sec:continuous}. The action of an one-parameter class of transformations $F$ on the local weights is studied, for which the gauge coupling of Z(2) Higgs-gauge theory in two dimensions changes continuously, compensated by a gradual appearance of an Ising interaction at links. This gives an example of how continuous transformations can be dealt with in our algebraic formalism, and leads to a class of symmetries among a family of generalised Ising-gauge models which we will describe in Section \ref{sec:continuous}. These generalised models are equivalent to the Higgs-gauge theory, as will be shown.

\subsection{Two-dimensional dualities} \label{subsec:2dweg}

\paragraph*{Pure gauge theory.}

We first consider, mostly for didactic reasons, the case of Z(2) pure gauge theory on two-dimensional triangulations. It is an well known fact that in two dimensions lattice gauge theories are soluble in general \cite{witten}, and so this case is to be considered as a first simple application of the transformation formalism. We show how an explicit solution of the model can be found for arbitrary triangulations.

The Z(2) pure gauge theory is described by the coproduct given in Eq. \ref{eq:cop-pure}. As the lattice is composed of triangular faces we let $x$ and $\beta$ satisfy the relations of Eq. \ref{eq:3f-fix}. In this case the algebra $\mathcal{H}$ reduces to
\begin{equation}
\begin{array}{ccc}
\phi_0 \cdot \phi_0 = \phi_1 \cdot \phi_1 = \rho^{-1} \cosh(x) \; \phi_0 + \rho^{-1} \sinh(x) \; \phi_1 \, \textrm{,} \\
\phi_0 \cdot \phi_1 = \phi_1 \cdot \phi_0 = \rho^{-1} \sinh(x) \; \phi_0 + \rho^{-1} \cosh(x) \; \phi_1 \, \textrm{.} \\
\Delta(\phi_0)  = \phi_0 \otimes \phi_0 \textrm{,} \qquad \Delta(\phi_1) = \phi_a \otimes \phi_1 \, \textrm{,} \label{eq:H-pure-2d}
\end{array}
\end{equation}
with $e^{-2 \beta_g} = \tanh(3x)$ and $\rho^6 = 2 \sinh(6x)$. The face and link local weights are the coefficients of the tensors $M_{abc}$ and $\Delta^{ab}$, respectively, since all faces have three sides and all limks are shared by exactly two faces. Now consider the change of basis given by
 \begin{equation}
E = \pmatrix{\frac{1}{2} \rho e^{-x} & \frac{1}{2} \rho e^{-x} \cr 
	\frac{1}{2} \rho e^{x} & -\frac{1}{2} \rho e^{x}} \textrm{,} \label{eq:E-latt-dual}
\end{equation}
where $\xi_a=E_a^b \, \phi_b$. In the basis $\mathcal{B}^\prime = \{\xi_0,\xi_1\}$, the local weights read
\begin{equation}
\begin{array}{c}
\Delta^{\prime \; ab} = 2 \rho^{-2} \exp[(-1)^a 2x] \delta_{ab} \, \textrm{,}\\
 M_{abc}^\prime = \delta_{ac} \, \delta_{bc} \, \textrm{.} \label{eq:duality1}
\end{array}
\end{equation}
The interpretation of these transformed weights is the following. The face weights are nonzero only if all spins on the face boundary are equal. Thus the new model has spin variables $\sigma_x$ at faces. For a link connecting two neighboring faces with spins $\sigma_x$ and $\sigma_y$ there corresponds a local weight $\Delta^{\prime \; xy}$. From Eq. \ref{eq:duality1} this weight vanishes if $x \neq y$. Thus a spin configuration has non-vanishing global weight only if, at all links over the lattice, only faces with equal configurations are glued. This is possible only if all faces of the lattice have the same configuration\footnote{The lattice is supposed to be connected, not composed of disjoint pieces.}. Thus only two global configurations contribute to the partition function, those for which all faces have spin $+1$ or $-1$. Therefore, the partition function for a two-dimensional triangulation $L$ is
\begin{equation}
\begin{array}{ccc}
Z_g^{(2)}(L,\beta_g)|_{h=0} &=& (2 \rho^{-2})^{N_l} (e^{2 N_l x} + e^{- 2 N_l x} ) \\
&=&  2^{N_l +1}\rho^{-2 N_l} \cosh(2 N_l x) \, \textrm{,} \label{eq:2dpure-sol}
\end{array}
\end{equation}
where $N_l$ is the number of links of $L$. This expression can be rewritten in terms of $\beta$ as
\begin{equation} 
Z_g^{(2)}(L,\beta_g)|_{h=0} = 2^{3/2 N_f} \cosh^{N_f} \beta \, (1 + \tanh^{N_f} \beta) \, \textrm{,}
\end{equation}
where we have used that, for any triangulation, $N_l = 3/2 N_f$, and that $\exp^{-2\beta}=\tanh 6x \Rightarrow \exp^{-6x}=\tanh \beta$. The partition function depends only on the area of the surface, that is, on the number $N_f$ of triangular faces.

If a square lattice was used instead of a triangulation, then all faces would have four spin variables on its boundary, and the local face weights would be the coefficients of the tensor $M_{abcd}$. In this case the parameters $\beta$ and $x$ must satisfy $e^{-2 \beta_g} = \tanh(4x)$, $\rho^8 = 2 \sinh(8x)$. The transformation $E$ applied to this algebra gives the same model as in the case of triangulations. Repeating the arguments already used, we find that the partition function of the Z(2) pure gauge theory for a square lattice $L$ is given by
\begin{equation} 
Z_g^{(2)}(L,\beta_g)|_{h=0} = 2^{2 N_f} \cosh^{N_f} \beta \, (1 + \tanh^{N_f} \beta) \, \textrm{.}
\end{equation}
Again $Z$ depends only on the area of the surface, but the dependence is slightly different from the one occurring for triangulations. We conclude that the partition function of the theory depends on the chosen lattice and on its area, a result which agrees with those of \cite{witten,quasitop}.

\paragraph*{Higgs-gauge theory.}

Let us consider the Z(2) Higgs-gauge theory on two-dimensional triangulations. The algebra which describes this model is the same as in Eq. \ref{eq:H-pure-2d}, but with the coproduct of Eq. \ref{eq:cop-2d}, that is,
\begin{equation}
\begin{array}{ccc}
\phi_0 \cdot \phi_0 = \phi_1 \cdot \phi_1 = \rho^{-1} \cosh(x) \; \phi_0 + \rho^{-1} \sinh(x) \; \phi_1 \, \textrm{,} \\
\phi_0 \cdot \phi_1 = \phi_1 \cdot \phi_0 = \rho^{-1} \sinh(x) \; \phi_0 + \rho^{-1} \cosh(x) \; \phi_1 \, \textrm{,} \\
\Delta(\phi_0)  = e^{h/2} \phi_0 \otimes \phi_0 \textrm{,} \qquad \Delta(\phi_1) = e^{-h/2}  \phi_a \otimes \phi_1 \, \textrm{,}
\end{array}
\end{equation}
with $e^{-2 \beta_g} = \tanh(3x)$ and $\rho^6 = 2 \sinh(6x)$. The transformation $E$ defined in Eq. \ref{eq:E-latt-dual} now leads to the transformed local weights
\begin{equation}
\begin{array}{ccc}
\Delta^{\prime \; 00}  =  2 \rho^{-2} e^{2x} \cosh h \, \textrm{,} \\
\Delta^{\prime \; 11}  =  2 \rho^{-2} e^{-2x} \cosh h \, \textrm{,}\\
\Delta^{\prime \; 01} = \Delta^{\prime \; 10} =  2 \rho^{-2} \sinh h \, \textrm{,}\\
M_{abc}^\prime = \delta_{ac} \, \delta_{bc} \, \textrm{.}
\end{array}
\end{equation}
These weights have a simple interpretation after some rearranging of the factors appearing in the above formulae is performed. First note that each link weight $\Delta^{\prime \; ab}$ contains a factor $\exp[(-1)^a x]\exp[(-1)^b x]$. The indices $a$ and $b$ describe the spins at the faces connected by the link. We let the factors $\exp[(-1)^a x]$ and $\exp[(-1)^b x]$ be absorbed by the corresponding face weights, i.e., we apply a rescaling with parameters $s_0=e^x$ and $s_1=e^{-x}$. Besides, a fixed factor $2 \rho^{-2} \sqrt{\sinh(h) \cosh(h)}$ is removed from all link weights. These factors will be later reinserted in the partition function. Then the local weights reduce to
\begin{equation}
\begin{array}{c}
\Delta^{\prime \; 00} =  \Delta^{\prime \; 11} = \sqrt{\coth h} \, \textrm{,} \\
\Delta^{\prime \; 01} =  \Delta^{\prime \; 10} = \sqrt{\tanh h} \, \textrm{,} \\
 M_{000}^\prime = e^{3x}, \qquad M_{111}^\prime = e^{-3x} \, \textrm{.} \label{eq:duality2}
\end{array}
\end{equation}
All the remaining $M_{abc}^\prime$ are null. These weights describe an Ising model on the dual lattice $L^\star$. Recall that for any face in $L$ there corresponds a vertice in $L^\star$. The weights $M_{abc}^\prime$ in Eq. \ref{eq:duality2} describe a model with spins at faces, so that on the dual lattice $L^\star$ we have spins at vertices. Let us write them as $\sigma_v=(-1)^v$, $v=0,1$. According to Eq. \ref{eq:duality2}, the dual vertice weights are $W(v)=\exp(\sigma_v \, 3x)$. Furthermore, for each link $l$ in $L$ there corresponds a link $l^\star$ in the dual lattice $L^\star$. The local weight at it is given by $\Delta^{\prime \, ab}$, where the indices $a,b$ describe the spins at the two ends of $l^\star$. Now if we write $\sqrt{\tanh(h)}=e^{-\beta_I^\star}$, then these link weights reduce to $W(l)=\exp(\beta_I^\star \sigma_a \, \sigma_b)$. And if we set $3x=m^\star$, then the vertice weights read $W(v)=\exp(\sigma_v \, m^\star)$. These are exactly the local weights for an Ising model at inverse temperature $\beta_I^\star$ in the presence of an external magnetic field $m^\star$ (see Eq. \ref{eq:weights-ising}).

Up to the removal of a factor $2 \rho^{-2} \sqrt{\sinh(h) \cosh(h)}$ from each link weight, we have found that the application of the transformation $E$ to the original Z(2) Higgs-gauge theory has made it an Ising model. Thus, up to such factors, the partition functions of the models are numerically equal. That is,
\begin{equation}
Z_g^{(2)}(L,\beta_g,h) = \left[2 \rho^{-2} \sqrt{\sinh h \cosh h}\right]^{N_l} Z_I^{(2)}(L^\star,	 \beta_I^\star, m^\star) \, \textrm{,} \label{eq:ZZ-2d}
\end{equation}
where $N_l$ is the number of links in $L$. The coupling constants are related by
\begin{equation}
\begin{array}{c}
e^{-2 \beta_I^\star} = \tanh h \, \textrm{,} \\
e^{-2 \beta_g} = \tanh m^\star \, \textrm{,} \label{eq:wegner2d-mag}
\end{array}
\end{equation}
where we have used Eq. \ref{eq:3f-fix} to write $x$ in terms of $\beta_g$. This is our version of the Kramers-Wannier bidimensional duality relation in the presence of an external magnetic field. These relations are valid when a triangulation is considered, instead of a square lattice.  The relations among the coupling constants is the same as obtained by Wegner. For the partition function, we can rewrite Eq. \ref{eq:wegner2d-mag} in terms of the coupling constants as
\begin{equation}
\frac{2^{N_f} e^{-\beta_g N_f}}{(\cosh h)^{N_l}} \, Z_g^{(2)}(L,\beta_g,h) = \frac{2^{N_l^\star} e^{-\beta_I^\star N_l^\star} }{(\cosh m^\star)^{N_v^\star}} \,  Z_I^{(2)}(L^\star, \beta_I^\star, m^\star) \, \textrm{,} \label{eq:Z2-2d-symm}
\end{equation}
where $N_f,N_l$ are the number of faces and links in $L$, and $N_v^\star,N_l^\star$ are the number of vertices and links in $L^\star$. The formula in Eq. \ref{eq:Z2-2d-symm} allows one to relate partition functions in triangulations, and is valid for arbitrary surfaces glued with any boundary conditions.

Let us also consider the case of square lattices. In this case the parameters $\beta$ and $x$ must satisfy $e^{-2 \beta_g} = \tanh(4x)$, $\rho^8 = 2 \sinh(8x)$. Applying again the transformation $E$ of Eq. \ref{eq:E-latt-dual}, we find the same transformed link weights $\Delta^{\prime \, ab}$ obtained earlier for triangulations, and given in Eq. \ref{eq:duality2}. For the face weights we find that $M_{abcd}^\prime = \delta_{ad} \, \delta_{bd} \, \delta_{cd}$. Repeating the arguments we used for the case of triangulations, but now taking $m^\star=4x$, we are again led to Eqs. \ref{eq:ZZ-2d} and \ref{eq:wegner2d-mag}. Moreover, the relation between the partition functions can be rewritten exactly as in Eq. \ref{eq:Z2-2d-symm}, which means that the same duality relations hold between the Z(2) Higgs-gauge theory and the Ising model in an external magnetic field for the cases of two-dimensional triangulations and square lattices.

The results of this section are extensions of the Kramers-Wannier dualities to arbitrary triangulations. The dualities are known to hold for any lattice, but the explicit expressions relating the partition functions on specific lattices are not known in general. The case of cubic lattices were worked out in the original paper. Our formalism allowed us to rewrite the expression in a way that it is valid for general finite compact triangulations. The relation between the two partition functions do not depend heavily on topological properties, as can be seen from Eq. \ref{eq:Z2-2d-symm}. Only simple combinatorial factors are involved, namely, the number of links and faces in the lattice.

\subsection{Three-dimensional dualities} \label{subsec:3dweg}

\paragraph*{Pure gauge theory.}

Now let the lattice $L$ be a three-dimensional triangulation. First let us consider the Z(2) pure gauge theory. In this case the parameters $\rho$ and $x$ must satisfy $e^{-2 \beta_g} = \tanh(3x)$ and $\rho^6 = 2 \sinh(6x)$, and the coproduct is the one given in Eq. \ref{eq:cop-pure}, so that the algebra is exactly the same as in the bidimensional case (Eq. \ref{eq:H-pure-2d}). In fact, this algebra describes the pure gauge theory on triangulations of any dimensionality $d$. Then the transformation $E$ gives the same local weights of Eq. \ref{eq:duality1}, but now link weights with an arbitrarily high number of indices may appear, as the number of faces meeting at a link in a three-dimensional triangulation is not fixed. Evaluating the weights, we find that
\begin{equation}
\begin{array}{c}
\Delta^{\prime \; a_1 a_2 \cdots a_m} = 2 \rho^{-m} \left\{ \prod_{i=1}^m \exp[(-1)^{a_i} x] \right\} \delta_{(-1)^{a_1 + a_2 + \cdots a_m},1} \, \textrm{,} \\
 M_{abc}^\prime = \delta_{ac} \, \delta_{bc} \label{eq:duality4} \, \textrm{.}
\end{array}
\end{equation}
It follows that the spin configurations are situated at faces after the transformation, since the face weights $M_{abc}^\prime$ are zero unless all spins on the face boundary are equal. And for the link weights $\Delta^{\prime \; a_1 a_2 \cdots a_m}$, we have the following. If the product of all spins around a link is equal to $+1$, then the corresponding weight is $2 \rho^{-m} \prod_{i=1}^m \exp[(-1)^{a_i} mx]$. Otherwise, the weight is zero. We would like to show that this theory is equivalent to a three-dimensional Ising model.

Consider an Ising model on a three-dimensional lattice $L$. Then there are spin variables $\sigma_v$ at the vertices of $L$, and local weights $W_I(l)=\exp(\beta  \sigma_x \sigma_y)$ at the links $l$ of $L$, where $\sigma_x, \sigma_y$ are the spins at the two ends of $l$. Let $\{ \sigma_v \}$ denote a spin configuration on the lattice. We can define a related configuration $\{ \sigma_l \}$, in which the spin variables are situated at the links of $L$. We just let the spin $\sigma_l$ at a link $l$ be the product of the spins at the two ends of this link, i.e., $\sigma_l=\sigma_x \sigma_y$. This assignment has two important properties. First, if all vertice spins $\sigma_v$ are flipped, then the same configuration $\{ \sigma_l \}$ is obtained, and the Ising weights $W_I=\exp(\beta  \sigma_l)$ remain unchanged. Up to this symmetry, $\{ \sigma_v \}$ is completely determined by $\{ \sigma_l \}$. Second, for any $\{ \sigma_l \}$ obtained by this prescription, we have that $(-1)^{f_1 + f_2 + \cdots f_m}=1$, where the indices $f_i$ describe the link spins at the sides of a face $f$. So we can rewrite the Ising model partition function on $L$ as
\begin{equation}
Z_I^{(3)}(L,\beta) = 2 \sum_{ \{\sigma_l  \}} \prod_l \exp(\beta  \sigma_l) \prod_f \delta_{(-1)^{f_1 + f_2 + \cdots f_m},1} \, \textrm{.} \label{eq:Ising-reform}
\end{equation}

We want to compare Eqs. \ref{eq:duality4} and \ref{eq:Ising-reform}. But first we perform some rearranging of factors in the transformed weights of Eq. \ref{eq:duality4}. Note that in the link weights $\Delta^{\prime \; a_1 a_2 \cdots a_m}$ there is a factor $\exp[(-1)^{a_i} x]$ for each face with configuration $a_i$ inciding at it. We perform a rescaling transformation to transfer this factor to the face weight. In addition, we remove a common factor $2 \rho^{-m}$ from each link weight, which will be later inserted back directly in the partition function. Then the local weights reduce to
\begin{equation}
\begin{array}{c}
\Delta^{\prime \; a_1 a_2 \cdots a_m} = \delta_{(-1)^{a_1 + a_2 + \cdots a_m},1}  \, \textrm{,}\\
 M_{000}^\prime = e^{3x} \, \textrm{,} \qquad M_{111}^\prime = e^{-3x} \, \textrm{,} \label{eq:duality4b}
\end{array}
\end{equation}
all the remaining $M_{abc}^\prime$ being null. On the dual lattice, these weights correspond to a model with spins at links, link weights $W(l)=\exp(3x \sigma_l)$, and face weights $W(f)=\delta_{(-1)^{a_1 + a_2 + \cdots a_m},1}$, where the indices $a_i$ describe the spins at the sides of the face $f$. Comparing with Eq. \ref{eq:Ising-reform}, we see that these weights describe an Ising model at inverse temperature $\beta_I^\star=3x$. Therefore, we have found that the transformation $E$, when applied to the local weights of a Z(2) Higgs-gauge theory on a three-dimensional triangulations $L$, leads to an Ising model in the dual lattice $L^\star$. From Eq. \ref{eq:3f-fix}, and recalling that $\beta_I^\star=3x$, we can write for the coupling constants the duality relation
\begin{equation}
e^{-2 \beta_g} = \tanh \beta_I^\star \, \textrm{,}
\end{equation}
which agrees with the result of Wegner for cubic lattices. For the partition functions we find that
\begin{equation}
Z_g^{(3)}(L, \beta_g) = 2^{-1} 2^{N_l} \rho^{-3N_f} Z_I^{(3)}(L^\star, \beta_I^\star) \, \textrm{.}
\end{equation}
This expression can be rewritten in terms of the coupling constants as
\begin{equation}
2^{N_f} e^{- \beta_g N_f} Z_g^{(3)}(L, \beta_g) = \frac{2^{N_f^\star -1}}{(\cosh \beta_I^\star)^{N_l^\star}} Z_I^{(3)}(L^\star, \beta_I^\star) \, \textrm{.}
\end{equation}
This relation holds for partition functions evaluated on arbitrary finite three-dimensional triangulations $L$, of any finite size and topology. The above expression also holds for cubic lattices $L$, with the only difference that then one must take $\beta_I^\star = 4x$. The explicit form of this relation written in a manner which is valid for general triangulations is, as far as we know, not available in the literature, and gives our generalisation of the Kramers-Wannier dualities of Wegner already extensively studied in cubic lattices. The relation between the partition functions depend on the topology of the lattice only through simple combinatorial factors, namely, the number of faces and links of the lattice.

\paragraph*{Higgs-gauge theory.}

Let us now consider the case of the Higgs-gauge theory. Let $L$ be a cubic lattice. In this case the model is described algebraically by
\begin{equation}
\begin{array}{ccc}
\phi_0 \cdot \phi_0 = \phi_1 \cdot \phi_1 = \rho^{-1} \cosh(x) \; \phi_0 + \rho^{-1} \sinh(x) \; \phi_1 \, \textrm{,} \\
\phi_0 \cdot \phi_1 = \phi_1 \cdot \phi_0 = \rho^{-1} \sinh(x) \; \phi_0 + \rho^{-1} \cosh(x) \; \phi_1 \, \textrm{,} \\
\Delta(\phi_0)  = e^{h/4} \phi_0 \otimes \phi_0 \, \textrm{,} \qquad \Delta(\phi_1) = e^{-h/4}  \phi_a \otimes \phi_1 \, \textrm{,}
\end{array}
\end{equation}
with $e^{-2 \beta_g} = \tanh(4x)$ and $\rho^8 = 2 \sinh(8x)$. The action of the transformation $E$ now leads to the transformed weights
\begin{equation}
\begin{array}{ccc}
\Delta^{\prime \, a_1 a_2 a_3 a_4} = 2\rho^{-4} \left[ \prod_{i=1}^4 e^{(-1)^{a_i} x} \right] \left(\cosh h \,  \delta_{(-1)^{a_1+a_2+a_3+a_4},1} +  \sinh h \, \delta_{(-1)^{a_1+a_2+a_3+a_4},-1} \right) \textrm{,} \\
M_{abcd}^\prime = \delta_{ad} \, \delta_{bd} \, \delta_{cd} \, \textrm{.}
\end{array}
\end{equation}
The interpretation of these weights is the following. As the face weights vanish unless all spins on a face boundary are equal, then we have spin variables situated at faces, say $\sigma_f$. Four spins meet at each link. If their product is $1$, then the link weight is $2\rho^{-4}  \cosh h \prod_{i=1}^4 e^{(-1)^{a_i} x}$, and if it is $-1$, then the link weight is $2\rho^{-4} \sinh h \prod_{i=1}^4 e^{(-1)^{a_i} x}$. As in previous cases, we collect some factors to make the weights simpler. Note that there is a factor $(-1)^{a_i}$ for each spin inciding at a link. We transfer this factor to the corresponding face weight. In addition, we remove a common factor $2 \rho^{-4} \sqrt{\sinh h \cosh h}$ from each link weight, which we will later inserted back in the partition function. Then the weights reduce to
\begin{equation}
\begin{array}{ccc}
\Delta^{\prime \, a_1 a_2 a_3 a_4} = \left(\tanh^{-1/2} h \,  \delta_{(-1)^{a_1+a_2+a_3+a_4},1} +  \tanh^{1/2} h \, \delta_{(-1)^{a_1+a_2+a_3+a_4},-1} \right) \textrm{,} \\
M_{0000}^\prime = e^{4x} \, \textrm{,} \qquad M_{1111}^\prime = e^{-4x} \, \textrm{,}
\end{array}
\end{equation}
all the remaining face weights being zero. Putting $e^{-2 \beta_g^\star} = \tanh h$, and $h^\star = 4x$, this is the same as
\begin{equation}
\begin{array}{ccc}
\Delta^{\prime \, a b c d} = \exp \left[\beta_g^\star (-1)^{a+b+c+d}\right] \, \textrm{,} \\
M_{0000}^\prime = e^{h^\star} \, \textrm{,} \qquad M_{1111}^\prime = e^{-h^\star} \, \textrm{.} \label{eq: 3d-pure-dual}
\end{array}
\end{equation}
Now these weights have a direct interpretation on the dual lattice. In the lattice $L$, the spins are situated at faces. Thus in the dual lattice $L^\star$, they lie on links. According to Eq. \ref{eq: 3d-pure-dual}, there is a link weight $W_l = \exp({\sigma_{l^\star}h^\star})$ for each link with a spin $\sigma_{l^\star}$ at it. Furthermore, the dual face weights are given by the coefficients of $\Delta^{\prime \, a b c d}$, which describe a four-spin gauge interaction with coupling constant $\beta_g^\star$. That is, another Z(2) gauge-Higgs theory is obtained on the dual lattice. We can write for the the coupling constants the duality relations
\begin{equation}
\begin{array}{ccc}
e^{-2 \beta_g^\star} = \tanh h \, \textrm{,} \\
e^{-2 \beta_g} = \tanh h^\star \, \textrm{,}
\end{array}
\end{equation}
and for the partition functions we find that
\begin{equation}
Z_g^{(3)}(L,\beta_g,h) = \left[2 \rho^{-4} \sqrt{\sinh h \cosh h}\right]^{N_l} Z_g^{(3)}(L^\star, \beta_g^\star, h^\star) \, \textrm{.}
\end{equation}
This formula can be rewritten in terms of the coupling constants as
\begin{equation}
\frac{e^{- \beta_g N_l}}{(\cosh h)^{N_l}}Z_g^{(3)}(L,\beta_g,h) = \frac{e^{- \beta_g^\star N_f^\star}}{(\cosh h^\star)^{N_f^\star}} Z_g^{(3)}(L^\star, \beta_g^\star, h^\star) \, \textrm{.}
\end{equation}
This formula gives the self-duality relation between the partition functions of Z(2) Higgs-gauge theory on a cubic lattice and on its dual lattice. This is a known result. We have showed that such duality can be recasted as an algebraic symmetry, that is, interpreted as a change of basis in the algebra $\mathcal{H}$ which describes the model.

\paragraph*{Higgs-gauge theory with a variable Higgs coupling.}

As an example of a new duality among spin models which can be obtained in our formalism, let us consider the case of a Z(2) gauge-Higgs theory with a variable Higgs coupling on a three-dimensional triangulation $L$. Let there be a Higgs coupling $h_l=mh$ at each link $l$, where $m$ is the number of faces meeting at the link. The gauge coupling $\beta_g$ is held constant over the lattice. This theory is described algebraically by the coproduct
\begin{equation}
\Delta(\phi_a) = \exp[ (-1)^a h] ] \, \phi_a \otimes \phi_a \, \textrm{,}
\end{equation}
and the product for which $\rho$ and $x$ satisfy $e^{-2 \beta_g} = \tanh(3x)$ and $\rho^6 = 2 \sinh(6x)$. The link weights of this theory are just
\begin{equation}
\Delta^{a_1 a_2 \cdots a_m} = \delta_{a_1 a_m} \delta_{a_1 a_m} \cdots \delta_{a_1 a_m} \exp[(-1)^{a_m} mh] \, \textrm{,}
\end{equation}
as already discussed. The face weights are those of a Z(2) lattice gauge theory with gauge coupling $\beta_g$. The action of the transformation $E$ on these weights leads to the transformed local weights
\begin{equation}
\begin{array}{ccc}
\Delta^{\prime \, a_1 a_2 \cdots a_m} = 2\rho^{-m} \left[ \prod_{i=1}^m e^{(-1)^{a_i} x} \right] \left(\cosh mh \,  \delta_{(-1)^{\sum_{i=1}^m a_i},1} +  \sinh mh \, \delta_{(-1)^{{\sum_{i=1}^m a_i}},-1} \right) \textrm{,} \\
M_{abc}^\prime = \delta_{ab} \, \delta_{ac} \, \textrm{.}
\end{array}
\end{equation}
Let us interpret these weights. But first we perform some rearranging of factors, as in previous cases. Note that in each link weight, there is a factor $\rho^{-1} \exp[(-1)^{a_i}]$ for each face incident at the link. We transfer it to the corresponding face weight. A factor $\rho^{-3}$ shows up at every face, which we remove from them. These can be later multiplied directly to the partition function. We also remove a factor $2\sqrt{\cosh(mh) \sinh(mh)}$ from the link weights. Then the weights reduce to
\begin{equation}
\begin{array}{ccc}
\Delta^{\prime \, a_1 a_2 \cdots a_m} = \sqrt{\coth mh} \,  \delta_{(-1)^{\sum_{i=1}^m a_i},1} +  \sqrt{\tanh mh} \, \delta_{(-1)^{{\sum_{i=1}^m a_i}},-1} \, \textrm{,} \\
M_{000}^\prime = e^{3x}\, \textrm{,} \qquad M_{111}^\prime = e^{-3x} \, \textrm{.} \label{eq:modified-dual}
\end{array}
\end{equation}
All the remaining face weights are zero. If we define new coupling constants $\beta_g^\star$ and $h^\star$ by
\begin{equation}
\begin{array}{c}
e^{-2\beta_g^\star(m)} = \tanh mh \, \textrm{,} \\
h^\star = 3x \, \textrm{,} \label{eq:mod-duality}
\end{array}
\end{equation}
then we can rewrite Eq. \ref{eq:modified-dual} as
\begin{equation}
\begin{array}{ccc}
\Delta^{\prime \, a_1 a_2 \cdots a_m} = \exp \left[ (-1)^{\sum_{i=1}^m a_i} \beta_g^\star(m) \right] \textrm{,} \\
M_{000}^\prime = e^{h^\star}\, \textrm{,} \qquad M_{111}^\prime = e^{-h^\star} \, \textrm{.}
\end{array}
\end{equation}
On the dual lattice $L^\star$ these weights have the following interpretation. There are spin variables $\sigma_{l^\star}$ at the links of $L^\star$, since in the original lattice they lie on faces. There corresponds to each link a local weight given by $\exp(h^\star \sigma_l)$, which describes a Higgs term. The face weights on $L^\star$ are the coefficients of $\Delta^{\prime \, a_1 a_2 \cdots a_m}$. These describe a gauge theory with a coupling constant $\beta_g^\star(m)$, whose strength is variable, and fixed by the duality relation of Eq. \ref{eq:mod-duality}. The Higgs coupling $h^\star$ does not vary. Furthermore, there is an extra weight at the dual faces yet, corresponding to the factors $w_d = 2 \sqrt{\cosh(mh) \sinh(mh)}$ that we have summarily removed from the face weights of $L$. These weights are of a new nature, and cannot be interpreted as arising from gauge or Ising interactions.

The strength of the gauge coupling decreases with the number of sides of the face. It means that in this model the gauge interactions are weaker in larger faces, where the spins are far from each other. If the lattice is regular, then all faces are equal, and we recover a Z(2) Higgs-gauge theory. In addition, the extra weighs $w_d = 2 \sqrt{\cosh(mh) \sinh(mh)}$ give a naive measure of the faces size. These weights depend only on the number $m$ of sides in the face, and increases with $m$. Besides, in the limit of $m>>1$, they reduce to $w_d=\exp(mh/2)$, which corresponds to an energy term $mh/2$ in the action. Recalling that the number of triangles in a face with $m$ sides is proportional to $m$, we can understand this term as proportional to the face area.

\section{Continuous symmetries in two dimensions} \label{sec:continuous}
In this section we consider the action of another class of transformations of GL(2,$\mathbf{R}$) on the local weights of Z(2) lattice gauge theory, as another example of application of the transformation formalism. We consider pure and Higgs-gauge theory on two-dimensional triangulations. We define a two-parameter class of transformations
\begin{equation}
F = 	\pmatrix{ \mu \cosh y  & - \mu \sinh y \cr 
		- \mu \sinh y  &  \mu \cosh y} \label{eq:F}
	\textrm{,}
\end{equation}
and study its action on the local weights $\Delta^{ab}$ and $M_{abc}$ of these theories. In terms of the new basis vectors $\xi_a = F_a^b \phi_b$, the algebraic structure of $\mathcal{H}$ is given by
\begin{equation}
\begin{array}{ccc}
\xi_0 \cdot \xi_0 = &\xi_1 \cdot \xi_1 = & \mu \rho^{-1} \cosh(x-y) \; \xi_0 + \mu \rho^{-1} \sinh(x-y) \; \xi_1 \, \textrm{,} \\
\xi_0 \cdot \xi_1 = &\xi_1 \cdot \xi_0 = & \mu \rho^{-1} \sinh(x-y) \; \xi_0 + \mu \rho^{-1} \cosh(x-y) \; \xi_1 \, \textrm{.}
\end{array}
\end{equation} 
If we set $\rho^{\star \, -1}= \mu \rho^{-1}$ and $x^\star = x-y$, then this algebra has the form given in Eq. \ref{eq:alg}. Therefore, it describes the face weights of another Z(2) gauge theory, with a transformed gauge coupling $\beta_g^\star$, given by $e^{-2 \beta_g^\star} = \tanh 3x^\star$. As we are working with triangulations, the parameter $\mu$ must be chosen so that $\rho^{\star \, 6} = 2 \sinh 6x^\star$. It follows that we must set 
\begin{equation}
\mu = \left(\frac{\sinh 6x}{\sinh 6x^\star}\right)^{1/6} \textrm{.}
\end{equation}
Granted that this relation is satisfied, $F$ reduces to a one-parameter class of transformations on $\mathcal{H}$, for which the gauge coupling $\beta_g$ changes continuously. The action of the transformation on the link weights depends on the value of $h$, and we consider the cases $h=0$ and $h\neq0$ separately.

\paragraph*{Pure gauge theory.}

\begin{figure}
\center
\includegraphics[scale=.55]{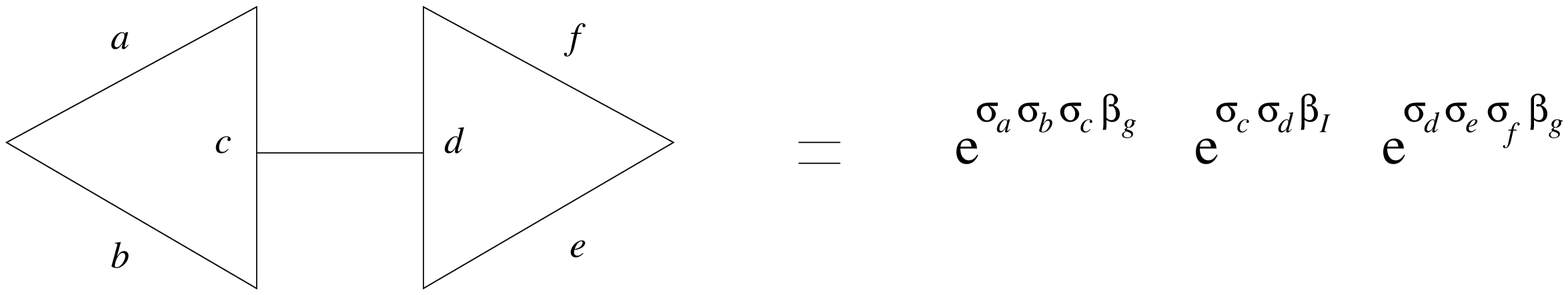}
\caption{Modified Z(2) pure gauge theory. There is a spin variable $\sigma_x=(-1)^x$ at each face side. The face weights are those of a Z(2) gauge theory. For each link there corresponds the local weight of an Ising model.}
\label{fig:modpure}
\end{figure}

The coproduct for Z(2) pure gauge theory is given by Eq. \ref{eq:cop-pure}. We find for the transformed link weights that
\begin{equation}
\begin{array}{ccc}
\Delta^{\prime \, 00} = \Delta^{\prime \, 11} = \mu^{-2} \cosh 2y \, \textrm{,}\\
\Delta^{\prime \, 01} = \Delta^{\prime \, 10} = \mu^{-2} \sinh 2y \, \textrm{.}
\end{array}
\end{equation}
The parameter $y$ must be positive in order that these weights are strictly positive. In this case, they describe an Ising interaction between the two spins incident at the link. This interpretation follows again from a simple rearranging of factors. Let us remove from all weights a common factor $\mu^{-2} \sqrt{\sinh 2y \cosh 2y}$, and define $\beta_I^\star$ by
\begin{equation}
e^{-2 \beta_I^\star} = \tanh 2y \, \textrm{.}
\end{equation}
Then the link weights reduce to $\Delta^{ab} = \exp[(-1)^{a+b} \beta_I^\star]$, which has the explicit form of an Ising interaction. In the new basis $\mathcal{B}^\prime = \left\{ \xi_0, \xi_1 \right\}$, we have the following transformed theory, which we call a modified gauge theory. The spin variables are assigned independently to the sides of the faces of $L$. There is a Z(2) gauge theory weight $M_{abc}(f) = \exp[(-1)^{a+b+c} \beta_g^\star]$ at each face $f$. Two spins meet at each link of the triangulation, and there corresponds to this an Ising weight $\Delta^{ab} = \exp[(-1)^{a+b} \beta_I^\star]$ (see Fig. \ref{fig:modpure}). The partition function of this modified theory is the sum over configurations
\begin{equation}
Z_{mod}^{(2)}(L,\beta_g^\star, \beta_I^\star) = \sum_{\left\{\sigma_a \right\}} \prod_f M_{f_1 f_2 f_3} \prod_l \Delta^{l_1 l_2} \, \textrm{,} \label{eq:mod-def}
\end{equation}
where the indices $f_i$ and $l_j$ describe the spins at the boundary of the face $f$ and at the two sides of the link $l$, respectively. 

As the modified theory we have described was obtained from Z(2) pure gauge theory by the action of the transformation $F$ given in Eq. \ref{eq:F}, then the partition functions are related by
\begin{equation}
Z_g^{(2)}(L,\beta_g) = \left(\mu^{-2} \sqrt{\sinh 2y \cosh 2y}\right)^{N_l} Z_{mod}^{(2)}(L,\beta_g^\star, \beta_I^\star) \, \textrm{,}
\end{equation}
where $N_l$ is the number of links in the triangulation. For the coupling constants, we find
\begin{equation}
\tanh^3 \beta_I^\star \cdot \tanh^2 \beta_g^\star = \tanh^2 \beta_g \, \textrm{,} \label{eq:cont-pure}
\end{equation}
It follows from Eq. \ref{eq:cont-pure} that two modified Z(2) gauge theories with coupling constants $\beta_I^1, \beta_g^1$ and $\beta_I^2, \beta_g^2$ have the same partition function, up to a factor, if
\begin{equation}
\tanh^3 \beta_I^1 \cdot \tanh^2 \beta_g^1 = \tanh^3 \beta_I^2 \cdot \tanh^2 \beta_g^2  \, \textrm{.}
\end{equation}
This formula describes a family of continuous symmetries among such spin-gauge models, and gives an explicit example of how the transformation formalism works for continuous transformations.

\paragraph*{Higgs-gauge theory.}

\begin{figure}
\center
\includegraphics[scale=.55]{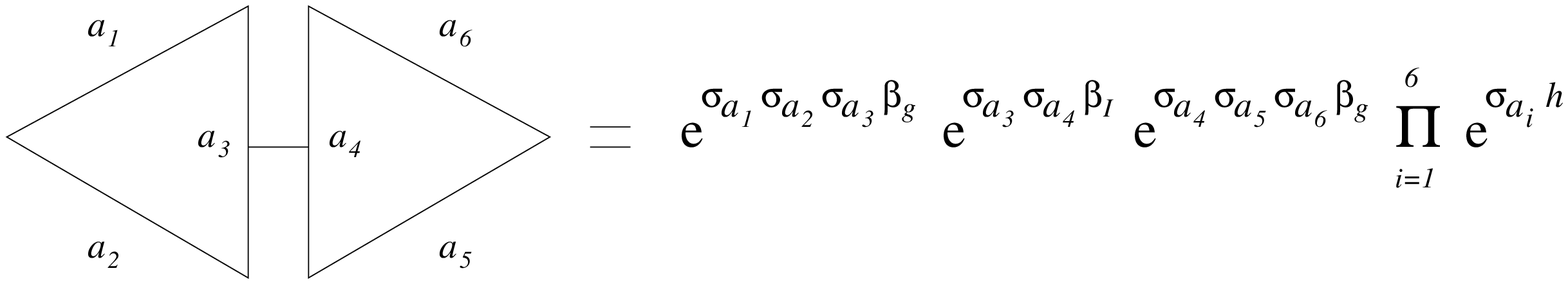}
\caption{Modified Z(2) gauge theory in the presence of a magnetic field $h$. There is a spin variable $\sigma_x=(-1)^x$ at each face side. The face weights are those of a Z(2) gauge theory. For each link there corresponds the local weight of an Ising model. For each spin configuration $\sigma_x$ there is a magnetic local weight $e^{\sigma_x h}$.}
\label{fig:modmag}
\end{figure}

Now let us see which modifications are introduced by the presence of a nonzero Higgs coupling $h$. In this case the coproduct is the one given in Eq. \ref{eq:cop-2d}, and the transformation $E$ leads to the transformed link weights
\begin{equation}
\begin{array}{ccc}
\Delta^{\prime \, 00} &= & \mu^{-2}(e^h \cosh^2 y + e^{-h} \sinh^2 y) \, \textrm{,} \\
\Delta^{\prime \, 01} = \Delta^{\prime \, 10} & = & \mu^{-2}\, (e^h + e^{-h}) \sinh y \cosh y  \, \textrm{,} \\
\Delta^{\prime \, 11} &= & \mu^{-2}(e^h \sinh^2 y + e^{-h} \cosh^2 y) \, \textrm{.}
\end{array}
\end{equation}
To understand the model these weights describe, let us rewrite them in a simpler form. Introducing the new coupling constants $\beta_I^\star$ and $h^\star$ defined by
\begin{equation}
\begin{array}{ccc}
e^{4 \beta_I^\star} = \frac{e^{2h} + \tanh^2{y}}{1 + e^{2h} \tanh^2 y} \, \textrm{,} \\
e^{4 h^\star} = 1 + \frac{1}{\cosh^2 h \sinh^2 2y} \, \textrm{,} \label{eq:cont-duality}
\end{array}
\end{equation}
and performing some algebraic manipulation, we find that the local weights reduce to
\begin{equation}
\begin{array}{ccc}
\Delta^{00} = q e^{2 h^\star} e^{\beta_I^\star} \, \textrm{,} \\
\Delta^{01} = q e^{- \beta_I^\star} \, \textrm{,} \\
\Delta^{11} = q e^{- 2 h^\star} e^{\beta_I^\star} \, \textrm{,} \label{eq:doubled}
\end{array}
\end{equation}
where
\begin{equation}
q^4 = (1 + \cosh^2 h \sinh^2 2y)(\cosh^2 h \sinh^2 2y) \, \textrm{.}
\end{equation}
Now these weights have a simple lattice interpretation. They describe another kind of modified Z(2) gauge theory, similar to the one obtained for the case of the pure gauge theory. The spin variables are attached independently to the sides on the faces boundaries along the lattice. There is a face weight $M_{abc}(f) = \exp[(-1)^{a+b+c} \, \beta_g^\star]$ at each face $f$. This describes a gauge interaction. There is also an weight $\Delta^{ab}(l) = \exp[(-1)^{a+b} \, \beta_I^\star]$ for each link $l$, where $a$ and $b$ are the spins connected by the link. This corresponds to an Ising interaction between such spins. Finally, for every spin variable $\sigma_a$ at the boundary of a face, there is a magnetic local weight $W_a = \exp[(-1)^a h^\star]$. A factor $q$ was removed from each link. A summary of the model is given in Fig. \ref{fig:modmag}.

The partition function is the product of all local weights in the lattice, summed over all spin configurations. We denote it by $Z_{mod}^{(2)}(L,\beta_g^\star, \beta_I^\star, h^\star)$. As it is obtained with the application of a change of basis in the 2D Higgs-gauge theory, it follows that
\begin{equation}
Z_{mod}^{(2)}(L,\beta_g^\star, \beta_I^\star, h^\star) = q^{N_l} Z^{(2)}_g(L,\beta_g,h) \, \textrm{,} \label{eq:mod-mag}
\end{equation}
where the parameters $\beta_I^\star, h^\star, h, y$ must satisfy the relations of Eq. \ref{eq:cont-duality}, and $e^{-2\beta_g}=\tanh 3x$, $e^{-2\beta_g^\star}=\tanh 3(x-y)$. The expression in Eq. \ref{eq:mod-mag} gives a family of continuous symmetries among the spin-gauge models $Z_{mod}^{(2)}(L,\beta_g^\star, \beta_I^\star, h^\star)$. In these symmetries, there is no need to invoke the dual lattice. Instead, we work with a less rigid discrete space, in which the combinatorial structures of both $L$ and $L^\star$ show up naturally. For such discretisation, the models related by the symmetry transformations live on the same lattice. A doubling of variables occurs, as we do not assign a single spin to each link, but one for each face meeting at the link. The models obtained are equivalent to the Z(2) Higgs-gauge theory, which in turn is dual to the bidimensional Ising model in the presence of an external magnetic field. Whether these modified theories may help in understanding the unsolved 2D transverse Ising model, we leave here as an open question.

\section{Conclusion and final remarks}

In this paper we studied a general class of symmetries relating partition functions of a family of classical spin models which generalises the Ising model and Z(2) lattice gauge theory. We interpreted these theories algebraically in terms of the structure constants of a vector space $\mathcal{H}$ equipped with algebra and coalgebra structures, using a formalism reminiscent from topological quantum field theories, and proved that to any change of basis of $\mathcal{H}$ there is an associated symmetry between these models. The well known Kramers and Wannier dualities of \cite{wegner} were shown to be special cases of this formalism. We studied these dualities on two- and three-dimensional triangulations, extending results known for the case of square and cubic lattices. In addition, a new class of continuous symmetry transformations involving generalised Z(2) gauge theories in two dimensions was developed.

In two dimensions, we found the following results. It is well known that pure gauge theories in two dimensions are exactly soluble. As a simple application of our formalism, we have described a change of basis of $\mathcal{H}$ which trivialises the theory and yields a solution for the partition function which is valid for any triangulation. The duality of the 2D Higgs-gauge to the Ising model coupled to a magnetic field was studied on general triangulations. Explicit expressions for the relation between the partition functions on general lattices were given, in terms of simple combinatorial factors such as the number of links and faces in the lattice. We also proved the existence of new symmetries relating the Z(2) Higgs-gauge theory to a generalised model described by three coupling constants, which describe a plaquette interaction, a magnetic field term, and an additional Ising interaction. The new symmetry transformations describe continuous transformations of these coupling constants.

In three dimensions, we studied the duality from the Z(2) pure gauge theory to the 3D Ising model on finite triangulations. We derived explicit expressions for the coefficients which relate the partition functions, in a manner which is valid for general triangulations. The expression depends on the topology of the lattice only through simple combinatorial properties, namely, the number of faces and links of the lattice. We also considered a modified Z(2) Higgs-gauge theory on finite triangulations, for which the Higgs coupling $h_l$ is variable along the lattice, being proportional to the number $m$  of faces meeting at the link $l$, $h_l=m h$. This theory was found to be dual to an Ising model with a variable inverse temperature, coupled to a magnetic field, and to an additional field which describes an area-dependent energy term in the action.

The method we have used to develop these symmetries can be applied to any theory which admits an algebraic interpretation similar to the one given in \cite{reform} for the case of Z(2) pure gauge theory on triangulations. Whether this is the case of more realistic gauge theories is an interesting question. In particular, it is natural to conjecture that dualities in Z(N) gauge theories can be formulated in a similar fashion. The extension to continuous gauge groups also deserves investigation. The mathematical prescriptions we have used do not pose any obstruction for the application to the case of non-commutative gauge groups. Duality relations are difficult to be formulated in this case, and our prescriptions might be of help in the study of this subject.

\section*{Acknowledgments}

We would like to thank J. C. A. Barata for many helpful discussions. Thanks also to A. P. Balachandran for suggestions and criticisms made in all phases of this work. N.Y. acknowledges the Institute of Physics at USP for the kind hospitality.

\end{document}